\documentclass{article}

\usepackage{arxiv}

\usepackage[utf8]{inputenc} 
\usepackage[T1]{fontenc}    
\usepackage{hyperref}       
\usepackage{url}            
\usepackage{booktabs}       
\usepackage{amsfonts}       
\usepackage{nicefrac}       
\usepackage{microtype}      
\usepackage{lipsum}		
\usepackage{graphicx}
\usepackage{natbib}
\usepackage{doi}

\usepackage{amsmath}

\title{Emergence of dynamical networks in termites}

\author{ \href{https://orcid.org/0009-0001-3726-1891}{\includegraphics[scale=0.06]{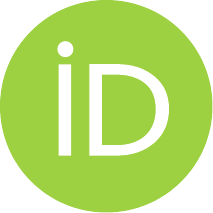}\hspace{1mm}Louis E.~Devers}\\
	Institut de Mathématiques de Toulouse (IMT)\\
	Université Paul Sabatier Toulouse III \\
	118 Rte de Narbonne, 31400 Toulouse \\
	\texttt{devers.louis@gmail.com} \\
	\And
	\href{https://orcid.org/0009-0002-2346-7845}{\includegraphics[scale=0.06]{orcid.pdf}\hspace{1mm}Perrine ~Bonavita} \\
	Centre de Recherche sur la Cognition Animale (CRCA-CBI)\\
	Université Paul Sabatier Toulouse III\\
	118 Rte de Narbonne, 31400 Toulouse \\
	\texttt{perrine.bonavita@univ-tlse3.fr} \\
    \And
	\href{https://orcid.org/0000-0003-3121-3001}{\includegraphics[scale=0.06]{orcid.pdf}\hspace{1mm}Christian ~Jost} \\
	Centre de Recherche sur la Cognition Animale (CRCA-CBI)\\
	Université Paul Sabatier Toulouse III\\
	118 Rte de Narbonne, 31400 Toulouse \\
	\texttt{christian.jost@univ-tlse3.fr} \\
}

\renewcommand{\shorttitle}{Emergence of dynamical networks in termites}

\hypersetup{
pdftitle={Emergence of dynamical networks in termites},
pdfsubject={nlin.AO},
pdfauthor={Louis E.~Devers, Perrine ~Bonavita, Christian ~Jost},
pdfkeywords={Dynamical Networks, Social Insects, Network Reconstruction, Termites, Biological Networks},
}

\begin{document}
\maketitle

\begin{abstract}
	Termites form complex dynamical trail networks from simple individual rules when exploring their environment. To help identify those simple rules, we reconstructed trail networks from time-lapse images of roaming termites. We quantified the trails' frequentations over time and compared them to the ones obtained by a null model. Arena borders were preferred in both simulated and observed data. Yet, the amplification phenomenon was higher with real termites, underlining the role of pheromones.
\end{abstract}


\section{Introduction}

In social insects, one can consider that the whole is more than the sum of its parts. Colony-level properties emerge from simple individual-based rules. For instance, in ants, it has been shown that the pheromones deposited by individuals allowed the colony to better exploit food sources \cite{deneubourg_collective_1989,deneubourg_probabilistic_1983,hughes_how_1990,dorigo_ant_1996}. 
Termites, similarly to ants, build nests, forage, form tunnel networks or even cultivate fungi\cite{bignell_biology_2011}. If pheromone trail emergence has been well studied in ants \cite{perna_individual_2012}, it is not the case of termites. Studies on termites mainly focused on the tunnelling network \cite{jmhasly_system_1999}, their dynamics \cite{mizumoto_complex_2020,jost_comparative_2012} and their nest architecture \cite{perna_topological_2008,perna_structure_2008,heyde_self-organized_2021}.
Unlike tunnelling behaviours, trail networks can be more difficult to observe. The movements of termites on surfaces without any building material, artificial galleries, or nest-oriented behaviours are not that well documented. Just like ants, termites' movements might be influenced by: other individuals \cite{paiva_scale-free_2021}, angles \cite{lee_path_2016,sim_direction_2017} and pheromones \cite{bignell_biology_2011,traniello_recruitment_1982,traniello_foraging_1989}. 

This paper aims to investigate individual behaviours that are sufficient and necessary to reproduce higher-level properties. In our case, we will focus on the trail network formed by freely roaming termites without any stimuli (nest, gallery, building material). Which are the individual rules reproducing such networks? and how to describe such networks?
We detail how we can extract a trail network of invisible pheromones through image processing. And we detail a method to follow the network's dynamical properties over time. To further explore this network, we developed a null model based on simple individual and voluntarily naive local rules. By comparing our observations to our null model, we can further understand how those networks are formed.

\section{Methods}
\subsection{Setup and species}

The experiment consists of filming termites roaming freely in a circular box and analysing the network they form in 20 minutes (Fig.\ref{fig:Method}). Termites will seach a shelter in such unfamiliar environment. The termites used: \textit{Procornitermes araujoi} measure about 5-6mm and originate from South America \cite{fouquet_construction_2011}. Experiments were run in 2012 by Christian Jost and Christine Lauzeral in Rio Claro, Brazil. The experiment was replicated 15 times. For each experiment, 106 \textit{Procornitermes araujoi} were extracted from the same nest on the university campus of UNESP Rio Claro. To maintain the polymorphism in natural populations, 100 of them were workers (smaller termites), and 6 of them were soldiers (bigger termites). Termites were contained in a 3cm diameter zone before being let free in arenas of 24 or 40cm diameter (respectively 6 and 9 replicates). Experiments were filmed at 25fps, and one picture in ten was kept (one every 0.4s). Thomas Colin segmented the termites into 3000 binary pictures for each experiment. 

\subsection{Network reconstruction}

\begin{figure}[h!]
    \centering
    \includegraphics[width=0.8\linewidth]{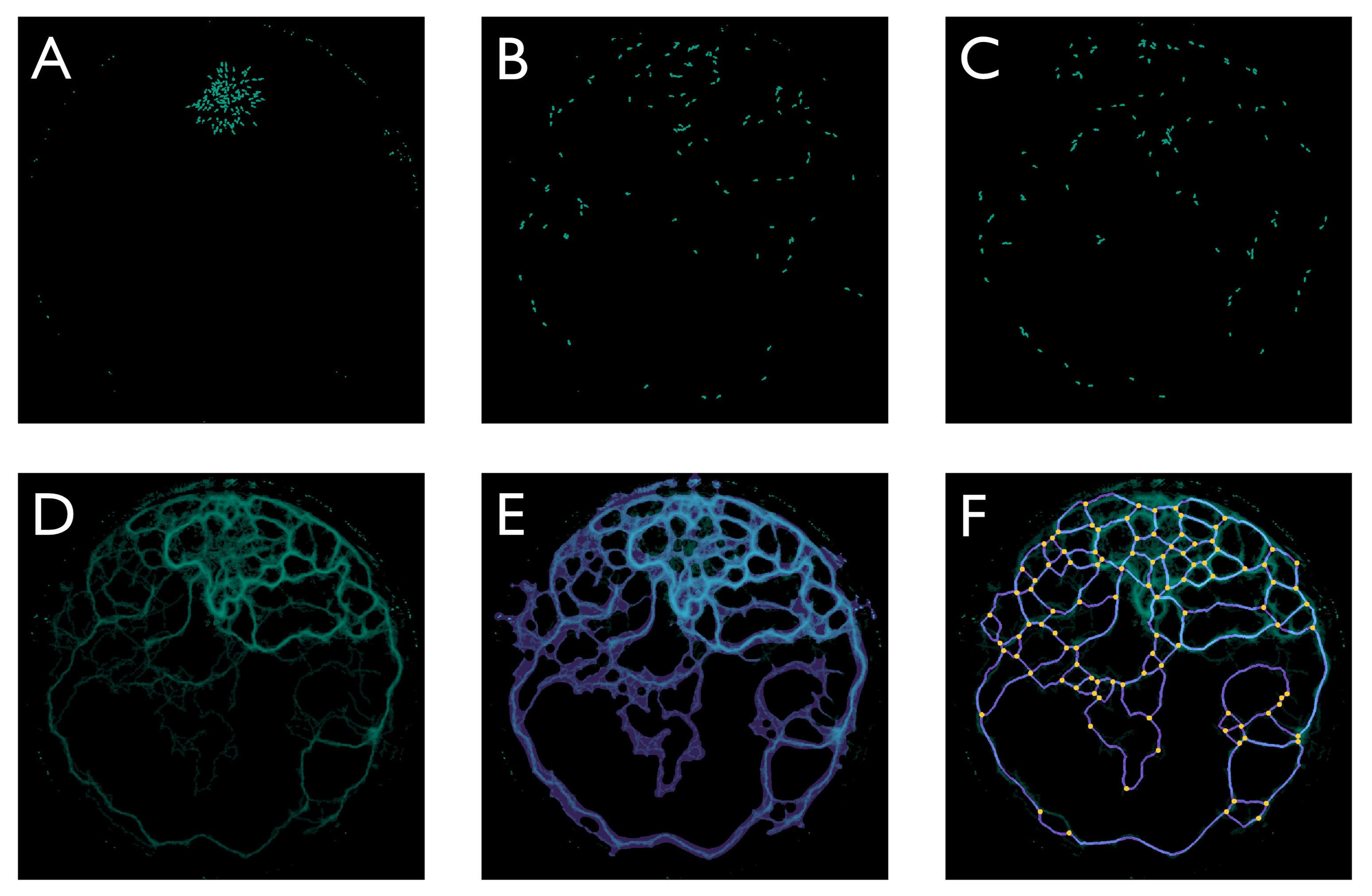}
    \caption{Method of network reconstruction from time-lapse images (arena diameter: 40cm). (A-C) are images of the binarised termites spreading and exploring the arena at time $t=0,5,10$ minutes. (D) is the cumulative image of moving termites' presence. (E) is the segmentation of the previous cumulated image (obtained using Frangi filters \cite{frangi_multiscale_1998}). (F) is the obtained network overlapped with the cumulated image for reference.}
    \label{fig:Method}
\end{figure}

For each experiment, we want to form a network from these segmented images and get the termite flow observed on each edge over time. We thus treated the images obtained using Matlab \cite{matlab_matlab_2016} (Fig.\ref{fig:Method}). 

We subtracted each frame from the previous one to obtain a binary mask of (only) moving termites. We then summed the binarized differentiated images before applying a log transformation to it (Fig.\ref{fig:Method}D) over the first 20 minutes (3000 pictures). The brighter the image, the more frequented it is. We can already observe that some paths are more frequented than others. To segment the network, we detected vessel-like objects using a Frangi filter \cite{frangi_multiscale_1998}. As it does not detect intersections, we obtained the whole network binary mask through additional morphological operations (Fig.\ref{fig:Method}E). We spurred this mask to form edges and split them by placing nodes on intersections and extremities (Fig.\ref{fig:Method}F). Some nodes were regrouped if too close to each other, thus forming nodes of degrees higher than 2 or 3. This process can be applied to any image (or stacked image) of a network, feel free to approach authors for more information.

We used the termites' binary masks to compute the dynamics of termite fraction on edges. If a termite is within the binary mask in Fig.\ref{fig:Method}E, its pixels get assigned to the nearest edge, thus giving non-directional data of all termite fraction on edges over time $N_{ij}(t)$. Notably, the sum of $N_{ij}(t)$ over all edges equals 1 for all time-steps.

\subsection{Null model}

There is little information about freely roaming insects network properties in the literature. Insects networks are usually studied in foraging (when they form networks around their nest) or nest-building context. To better compare the properties of this dynamical network we needed a null model \cite{buhl_growth_2006}. We propose here a freely roaming termite deterministic null model. The termites can move freely within all possible edges in the observed network (detected in the first 20 minutes of experiment). The model functions as follows: 

\begin{equation} \label{eq:1}
\vcenter{\hbox{\includegraphics[width=5cm,height=5cm]{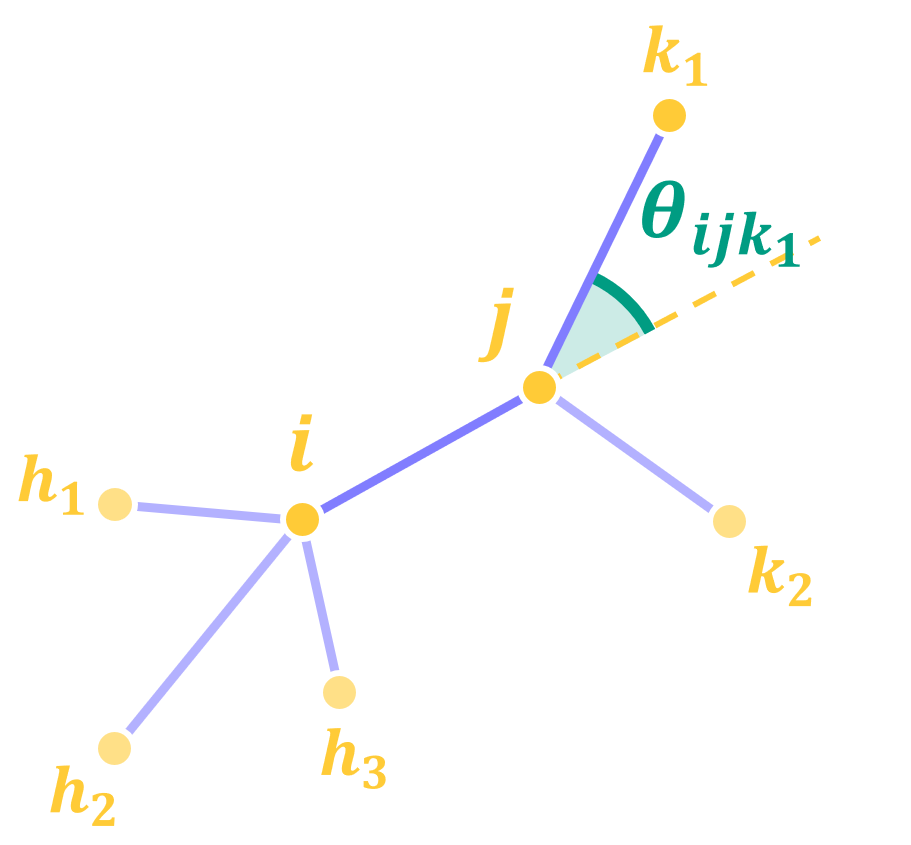}}}
\qquad\qquad
\begin{aligned}
    p_{ijk} = \frac{|\cos\left(\frac{\theta_{ijk}}{2}\right)|}{\sum_l |\cos\left(\frac{\theta_{ijl}}{2}\right)| }
\end{aligned}
\end{equation}{}

We noted $p_{ijk}$ the probability of joining the edge $jk$ (going from node $j$ to node $k$) for an agent present in the edge $ij$. The probability of joining $jk$ is computed as a ratio of a preference score over all the possible $jl$ edges accessible from node $j$. That preference score is computed as $|\cos\left(\frac{\theta_{ijk}}{2}\right)|$ where $\theta_{ijk}$ is the turning angle of a termite moving from edge $ij$ and edge $jk$. The preference score equals one if $i$, $j$ and $k$ are aligned ($\theta_{ijk}=0$), and 0 if going backwards (if $i=k, \theta_{ijk}=\pi\text{ or } -\pi$). In the illustration of Eq. (\ref{eq:1}), $\theta_{ijk_1} = \pi/6$ and $\theta_{ijk_2} = -\pi/3$. The preference scores $|\cos\left(\frac{\theta_{ijk}}{2}\right)|$ are respectively 0.9659 and 0.8660 for $k_1$ and $k_2$. Note that the preference score of going to $i$ from $ij$ is 0. Thus by Eq. (\ref{eq:1}): $p_{ijk_1} = 0.5273$ and $p_{ijk_2} =  0.4727$.

We can then write that $N_{ij}(t)$, the termite fraction on an edge $ij$ at time $t$, fluctuates as :

\begin{equation} \label{eq:2}
    \frac{dN_{ij} (t)}{dt} = v \times \left( \sum_h \frac{N_{hi}(t) p_{hij}}{L_{hi}} - \sum_k \frac{N_{ij}(t) p_{ijk}}{L_{ij}} \right) 
\end{equation}{}

In Eq. (\ref{eq:2}), $N_{ij} (t)$ is the termites fraction on the edge going from nodes $i$ to $j$ at time $t$. $v$ is a single termite velocity ($1 \text{ cm.s}^{-1}$). $N_{ij} (t)$ evolves positively with incoming termites coming from all possible nodes $h$, connected to $i$. The incoming flux is averaged to the termite fraction in $hi$ times the probability to join $ij$ from $hi$ ($p_{hij}$ in Eq. (\ref{eq:1})). The incoming flux must be divided by the length of said edge, $L_{hi}$, while the termite goes at a velocity $v$. Similar reasoning is made for leaving fluxes: $N_{ij} (t)$ evolves negatively with leaving termites going to all possible nodes $k$, connected to $j$. The leaving flux is averaged to the termite fraction in $ij$ times the probability to leave $ij$ to $jk$ ($p_{ijk}$ in Eq. (\ref{eq:1})). The leaving flux must be divided by the length of said edge, $L_{ij}$, while the termite goes at a velocity $v$.

To determine initial conditions, we identified the node $i$ closest to the termites' experimental release point. We evenly distributed termites in all out-going edges connected to node $i$. Similarly as the observed data, the sum of $N_{ij}(t)$ over all edges equals 1 for all time-steps.

These rules are simple, local, and only based on angle preferences. They roughly match termite angle preferences observed in tunnels \cite{lee_path_2016,sim_direction_2017}. Authors argue that the preference function can be any function returning one if edges are aligned ($\theta_{ijk}=0$) and returning 0 if going backward ($\theta_{ijk}= \pi \text{ or } -\pi$). 

\section{Results}

\subsection{Final states Networks}

To first describe the networks obtained, we will focus in this section on "final states networks". These networks include all the edges, and all the nodes extracted at t=20min. Each edge is associated with its termite fraction $N_{ij}(t)$ over the course of the experiment. For observed networks, we plotted the mean termite fraction considering all previous time-steps. For simulated networks, it consists of the final termite fraction (as the simulation reaches equilibrium corresponding to the mean termite fraction). We can thus compare simulated and observed termite fraction on edges.

\begin{figure}[h!]
    \centering
    \includegraphics[width=0.6\linewidth]{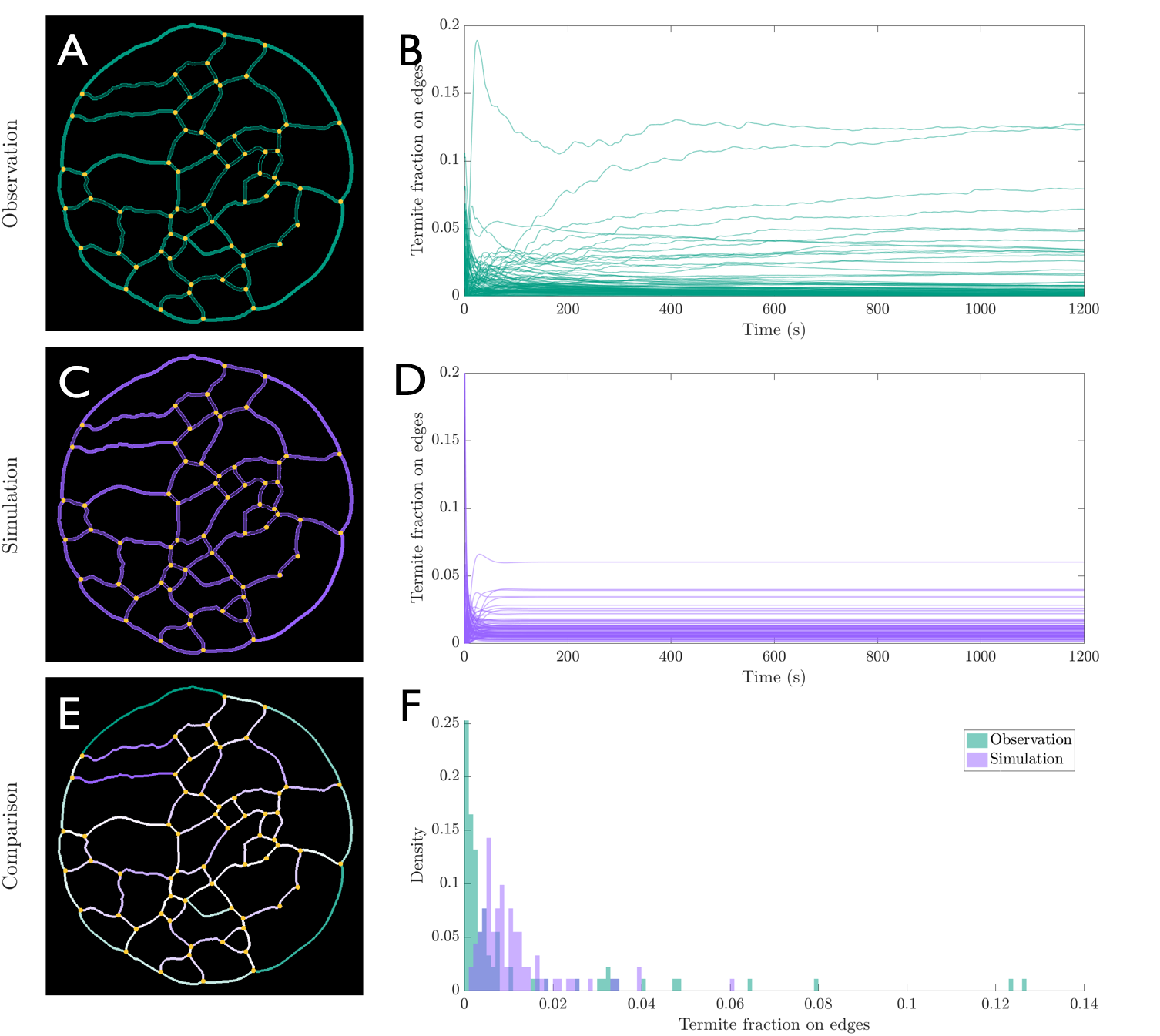}
    \caption{Termite fraction on edges over time in a single network (arena diameter: 24cm). (A) Extracted network and observed mean fraction. The filling of edges represents the fraction (dark to green for respectively low to high fraction). (B) Mean termite fraction over time. (C) Extracted network and simulated final termite fraction. The filling of edges represents the fraction (dark to purple for respectively low to high fraction). (D) Edge fraction over time (Eq. (\ref{eq:2})) (E) Difference of termite fraction on edges between observed and simulated data. Green edges are over-represented in the observed data, and purple edges are under-represented in observed data. White edges are equivalently dense in both. (F) Density distribution of observed mean termite fraction (green) and simulated final termite fraction (purple).}
    \label{fig:StaticNx}
\end{figure}

We can observe on Fig.\ref{fig:StaticNx} the obtained final network for both observation and simulation (Fig.\ref{fig:StaticNx}A and C respectively). The colour intensity corresponds to the final termite fraction on an edge. We can note that in the observed networks, termites are not uniformly distributed, especially compared to the simulated case. Indeed, some edges are highly preferred to others over time (Fig.\ref{fig:StaticNx}B). In the simulated case, termites do not exhibit strong preferences and rapidly reach a stable state (Fig.\ref{fig:StaticNx}D). Termite fraction on edges are not distributed in the same way (Fig.\ref{fig:StaticNx}F). 

To identify edges that are over-frequented in the observed network, we subtract observed and simulated termite densities (Fig.\ref{fig:StaticNx}E). Edges preferred by termites are in green, and edges preferred by the null model are in purple. Edges are white if densities are equivalent for the termites and the null model. In this specific network, edges on the border are over-frequented compared to our null model. Conjointly, most edges in the middle of the network are slightly preferred by the model.

\begin{figure}[h!]
    \centering
    \includegraphics[width=1\linewidth]{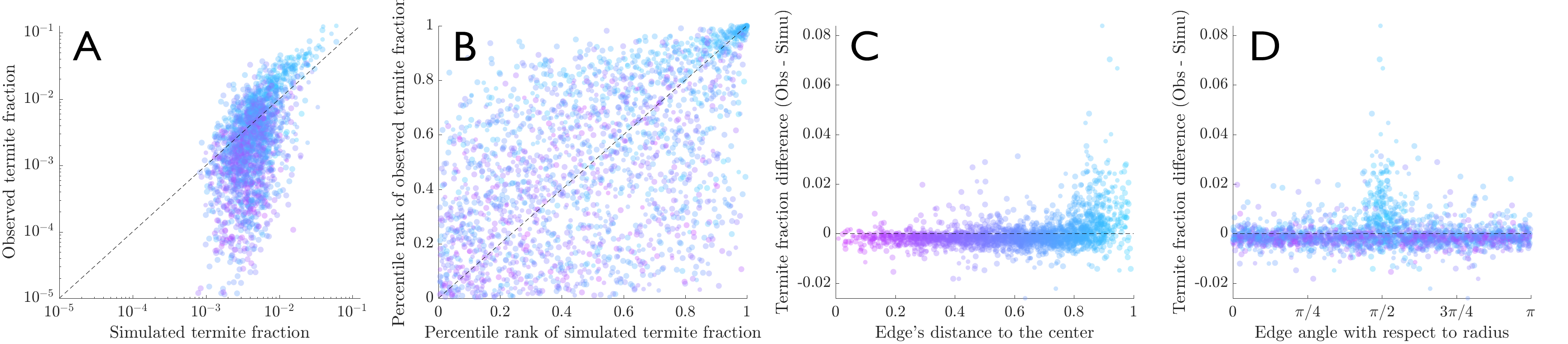}
    \caption{Comparison of simulated and observed termite fraction for each edge. (A) observed vs simulated termite densities (log-log scale). The diagonal dashed line visually supports testing for equality. (B) Percentile rankings of termite fraction (Simulated vs Observed) (C) Difference of termite fraction (Observed minus Simulated) as a function of edge position. Edge position was computed as its distance from the centre of the arena divided by the radius of the arena. The horizontal dashed line visually supports testing for equality. (D) Difference of termite fraction (Observed minus Simulated) as a function of edge orientation. Edge orientation was computed as its angle with the radius of the arena. The horizontal dashed line visually supports testing for equality. Colour is function to the edge's distance to the centre (pink for centre, bright blue for border edges) and marker size depends the diameter of the experimental arena.}
    \label{fig:StaticNxComparison}
\end{figure}

Can this observation be generalised to other networks ? In Fig.\ref{fig:StaticNxComparison} we compare observed simulated edges' termite fraction for all networks treated (12 out of 15). To visualise edges relative position in the arena, edges close to the border are represented in bright blue, while edges close to the centre of the area are represented in pink. In Fig.\ref{fig:StaticNxComparison}A, we plotted all observed edges' fraction against simulated ones. Over-represented edges compared to a null model are present over the dashed diagonal line (and respectively, under-represented ones bellow the line). Most edges are under-represented observations, meaning that termites prefer to focus on a few edges with a high activity. Bright blue points following the diagonal line in Fig.\ref{fig:StaticNxComparison}A show that edges located at the border of the arena are preferred in both models. However, the preference is way higher in actual termites' networks. This common preferences is also visible by plotting percentile rank of fraction of simulated vs observed edges (B). We note in the top-right corner that frequented edges are common in simulations and observation and correspond to border edges. However, other edges show few to no correspondence.

Which are the edges preferred in termites' networks ? From observation of Fig.\ref{fig:StaticNx}, we hypothesised that edges that are far from the middle of the arena are over-represented compared to a null model. We represented the difference of fraction (Observed - Simulated) against the edge position in the arena (Fig.\ref{fig:StaticNxComparison}C). Indeed, edges far from the centre of the arena (close to the border) are over-represented. We also plotted the difference of fraction against edge orientation with regard to the arena's radius (Fig.\ref{fig:StaticNxComparison}D). We note that edges perpendicular to the radius are over-represented compared to our null model. Both these observations support the fact that the network in Fig.\ref{fig:StaticNx} is representative of that phenomenon.   

\subsection{Dynamical Networks}

We showed in the previous section that the termite fraction on edges varies over time for both observed and null model networks. However, if an edge had a low termite fraction, meaning that a path was rarely frequented, it remained in the network. In this section, we propose a method to dynamically modify network topology as a function of edge fraction. Low fraction edges will be discarded and can be added back to the network later on. The structure of the network thus changes over time, and with it, its properties.

As seen in Fig.\ref{fig:StaticNx}B and D, the observed and simulated termite densities are not distributed in the same way. So, an absolute filter above which an edge is considered "active" will not suffice. To discriminate active and non-active edges, we propose a method inspired by social insects like ants and termites: pheromones. The amount of pheromones on a given edge increases with passing termites but decreases through evaporation at a constant rate $\mu$. Pheromones are usually key to understanding routing problems and path selection in social insects \cite{theraulaz_brief_1999,dorigo_ant_1996}. Here, we computed the amount of pheromones on each edge $Ph_{ij}$ for each time step as follows :

\begin{equation} \label{eq:3}
    \frac{dPh_{ij} (t)}{dt} = -\mu Ph_{ij} (t) + \frac{N_{ij}}{L_{ij}} 
\end{equation}{}

In Eq. (\ref{eq:3}), the concentration of pheromones $Ph_{ij}(t)$ on edge $ij$ evaporates at rate $\mu$. Previous work estimated the half life of \textit{Procornitermes araujoi} of being 16 minutes \cite{fouquet_construction_2011}. Implying a rate of evaporation of $\mu = 7.26 \times 10^{-4} \text{s}^{-1}$. The concentration of pheromones increases with the number of individuals present in edge $ij$. We need to divide by the length of the edge $L_{ij}$ to obtain concentrations of pheromones per cm.

From there, we conserved the edges with the higher amount of pheromones that totalled $p_{thresh}$ per cent of all the pheromones at time $t$. In our case,$p_thresh = 0.8$ meant that active edges were the biggest ones representing a total of 80\% of all pheromones. Such criterion allows easy comparison between the observed and simulated networks. 

\begin{figure}[h!]
    \centering
    \includegraphics[width=0.8\linewidth]{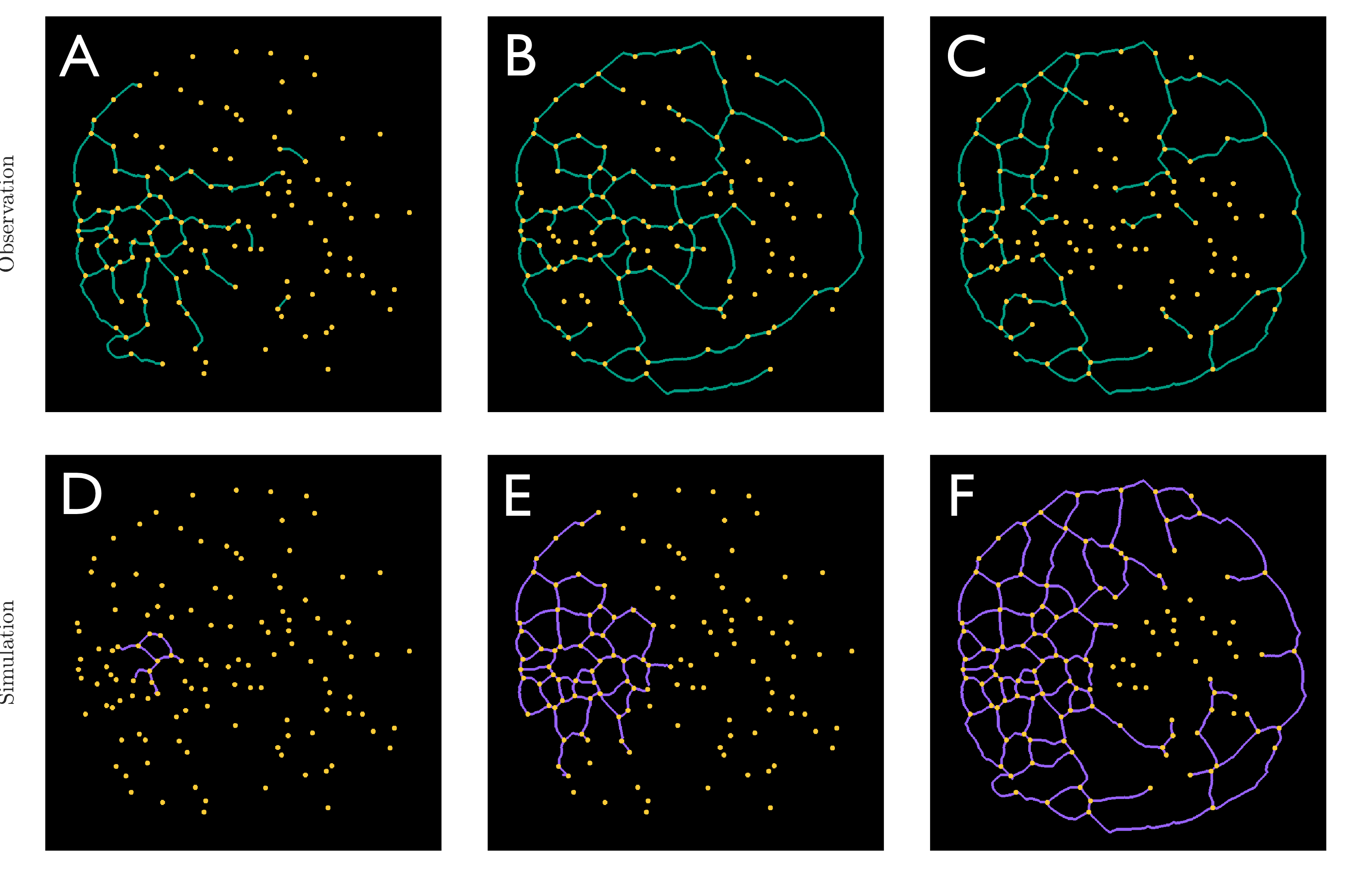}
    \caption{Examples of dynamical networks (arena diameter: 24cm). (A-C) Observed networks ($t=10,100,1000$ s). (D-F) Simulated networks ($t=10,100,1000$ s).}
    \label{fig:DynamicExample}
\end{figure}

In Fig.\ref{fig:DynamicExample}, is represented both observed (A-C) and simulated (D-F) networks over time ($t=10,100,1000s$). Concerning the observed network (A-C), we first observe a spread of the termites through the whole arena, followed by a selection of edges. The edges on the border are mainly selected. Concerning the networks simulated by our null model, we also observe a spread, but not followed by a drastic edge selection. However, border edges seem to be preferred as well. The main difference thus lies in the intensity of the filtering, rather than the edges being filtered.

\begin{figure}[h!]
    \centering
    \includegraphics[width=0.6\linewidth]{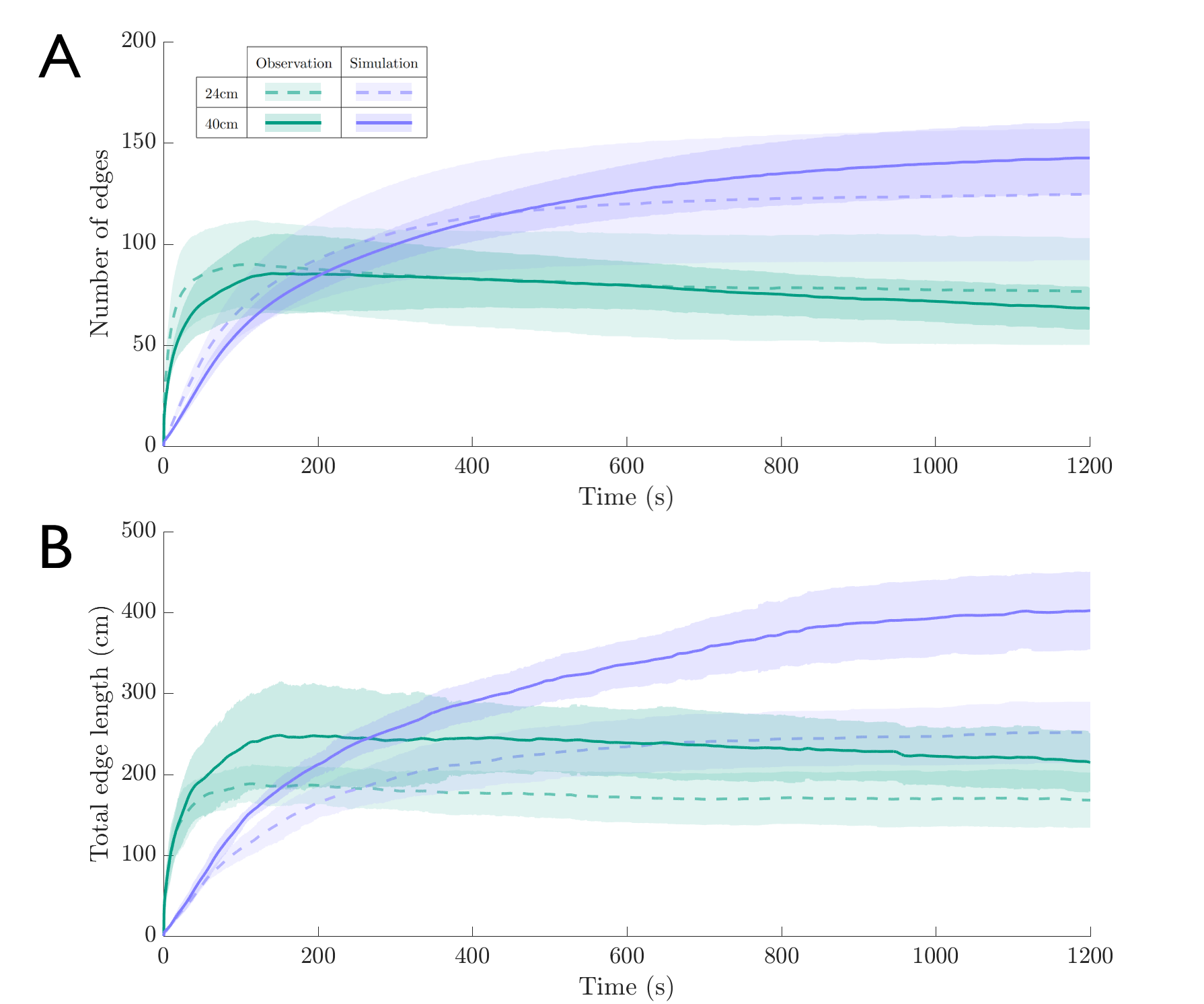}
    \caption{Mean length and number of edges (6 networks for each condition, confidence interval of one standard deviation). Observations in green, and simulations in purple. 24cm arenas with dashed lines and 40cm with solid lines.}
    \label{fig:DynamicComparison}
\end{figure}

The dynamics of the formed networks properties can be extensively studied. We propose here preliminary results concerning the total length of the networks and the number of conserved edges over time. Future work will be needed to focus on metrics like efficiency, robustness or meshedness for instance \cite{buhl_efficiency_2004,buhl_growth_2006,buhl_shape_2009}. In Fig.\ref{fig:DynamicComparison}, we represented (A) the total number of edges and (B) the total length of the network in cm over time. Both observed (green) and simulated (purple) networks are shown. We note that the number of edges and total length of the networks are different between the observed and simulated networks. However, we observe no differences in edge number and total length between 24 (light green) and 40cm (dark green) arenas. It could mean that 106 termites can only sustain a pheromone track of about 200cm independently of the arena's diameter.

\section{Discussion}

This paper aims to investigate individual behaviours that are sufficient and necessary to reproduce higher-level properties. In our case, we will focus on the trail network formed by freely roaming termites without any stimuli (nest, gallery, building material). Which are the individual rules reproducing such networks? and how to describe such networks?

In this paper, we observed termites forming networks while exploring a circular arena. We extracted its nodes and edges using image processing and Frangi filter \cite{frangi_multiscale_1998}. We measured the termite fraction of each edge over time, and underlined a preference for the border of the arena. Thigmotaxis, the preferences of animals for borders and contacts, is well known, especially in stressful situations \cite{schone_spatial_1984,creed_interpreting_1990,casellas_individual_2008}. So to assess whether that preference was due to the geometry of the arena, we established a null model simulating termites movements based on turning angles in the existing network. Our null model managed to explain the preference for the borders, without explicitly implementing it. However, the intensity of termites' edge selection was not reproduced. The individual rules we implemented in our null model were not sufficient to reproduce such collective behaviour. 

In our null model, agents prefer lower turning angles. Additionally, one can expect that pheromones drive an important role in the turning decisions \cite{perna_individual_2012}. Our model is missing the amplification some edge benefits, and pheromones play a key role in the amplification of an individual decision to a collective one \cite{dussutour_amplification_2005}. We also showed that the termites' networks total lengths stabilise around 200cm independently of the size of the arena. This fact supports the hypothesis of pheromone trails, as 106 individuals may only sustain a 200cm long pheromone track (considering evaporation rates). 
The future work should focus on improving the null model with a turning preference based on both angle and pheromone quantity. This next step will be straightforward from our data, as we already implemented pheromones in our model to discriminate between active and inactive edges. 
Our work would benefit from more pertinent network metrics especially suitable for planar network efficiency. New metrics will allow us to better differentiate our null models from observed collective behaviours over time. The future work should also focus on the survival analysis of edge activation depending on their location, branching, or orientation for instance \cite{vernet_study_2023}. 

\section{Acknowledgement}
We would like to kindly thank Olivier Giraud, Clément Sire, Guy Theraulaz and Ramon Escobedo Martinez, for their help with the aRxiv endorsment. We would like to thank Thomas Colin for segmenting the termites images. We would also like to thank the members of the Collective Animal Behaviour team in the CRCA-CBI of Toulouse for their time and their constructive feedback. A special thanks goes to Vincent Fourcassié for allowing me to pursue this project alongside my PhD thesis. 

\clearpage

\bibliographystyle{unsrtnat}
\bibliography{references}  






\end{document}